\documentclass[twocolumn]{aastex61}

\usepackage{my_astro}
\usepackage{courier}
\usepackage{amsmath}
\usepackage[super]{nth}

\newcommand{\JOIII}{$J_\mathrm{O_3}$}
\newcommand{\JCHIV}{$J_\mathrm{CH_4}$}
\newcommand{\madrel}{MAD$_\mathrm{rel}$}

\begin{document}

\title{HAZMAT. IV. Flares and Superflares on Young M Stars in the Far Ultraviolet\footnote{Based on observations made with the NASA/ESA Hubble Space Telescope, obtained from the data archive at the Space Telescope Science Institute. STScI is operated by the Association of Universities for Research in Astronomy, Inc. under NASA contract NAS 5-26555.}}

\correspondingauthor{R. O. Parke Loyd}
\email{parke@asu.edu}

\author{R. O. Parke Loyd}
\affiliation{School of Earth and Space Exploration, Arizona State University, Tempe, AZ 85287}

\author{Evgenya L. Shkolnik}
\affiliation{School of Earth and Space Exploration, Arizona State University, Tempe, AZ 85287}

\author{Adam C. Schneider}
\affiliation{School of Earth and Space Exploration, Arizona State University, Tempe, AZ 85287}

\author{Travis S. Barman}
\affiliation{Lunar and Planetary Laboratory, University of Arizona, Tucson, AZ 85721 USA}

\author{Victoria S. Meadows}
\affiliation{NASA Astrobiology Institute Alternative Earths and Virtual Planetary Laboratory Teams}
\affiliation{Department of Astronomy, University of Washington, Seattle, WA, USA}

\author{Isabella Pagano}
\affiliation{INAF - Osservatorio Astrofisico di Catania, Via S. Sofia 78, 95123, Catania, Italy}

\author{Sarah Peacock}
\affiliation{Lunar and Planetary Laboratory, University of Arizona, Tucson, AZ 85721 USA}

\accepted{2018 September 11}
\submitjournal{ApJ}

\begin{abstract}
M stars are powerful emitters of far-ultraviolet light.
Over long timescales, a significant, possibly dominant, fraction of this emission is produced by stellar flares.
Characterizing this emission is critical to understanding the atmospheres of the stars producing it and the atmospheric evolution of the orbiting planets subjected to it.
Ultraviolet emission is known to be elevated for several hundred million years after M stars form.
Whether or not the same is true of ultraviolet flare activity is a key concern for the evolution of exoplanet atmospheres.
\textit{Hubble Space Telescope (HST)} observations by the HAZMAT program (HAbitable Zones and M dwarf Activity across Time) detected 18 flares on young (40~Myr) early M stars in the Tucana-Horologium association over 10~h of observations, ten having energy $>10^{30}$ erg.
These imply that flares on young M stars are 100--1000$\times$ more energetic than those occurring at the same rate on ``inactive,'' field age M dwarfs.
However, when energies are normalized by quiescent emission, there is no statistical difference between the young and field age samples.
The most energetic flare observed, dubbed the ``Hazflare,'' emitted an energy of $10^{32.1}$~erg in the FUV, 30$\times$ more energetic than any stellar flare previously observed in the FUV with \textit{HST's} COS or STIS spectrographs.
It was accompanied by $15,500\pm400$~K blackbody emission bright enough to designate it as a superflare ($E>10^{33}$~erg), with an estimated bolometric energy of $10^{33.6_{-0.2}^{+0.1}}$~erg.
This blackbody emitted 18$^{+2}_{-1}$\% of its flux in the FUV (912--1700~\AA) where molecules are generally most sensitive to photolysis.
Such hot superflares in young, early M stars could play an important role in the evolution of nascent planetary atmospheres.
\end{abstract}


\section{Introduction}

M stars are extremely relevant to the quest for an understanding of the diversity, evolution, and biological potential of exoplanets.
They dominate planetary systems by number and their small masses and radii make their planets comparatively easy to detect and characterize (see \citealt{shields16} for a recent review).
These stars are known for their vigorous flaring, with flares
contributing a significant, potentially dominant portion of the far-ultraviolet (FUV) light they emit (\citealt{loyd18}; hereafter L18).
This emission has important consequences for planetary atmospheres.
FUV emission photolyzes molecules, perturbing the thermochemical equilibrium these atmospheres would otherwise achieve (e.g., \citealt{hu12, miguel14}).
Extreme ultraviolet (EUV) emission, formed alongside the FUV in the stellar upper atmosphere, powers thermal atmospheric escape (e.g., \citealt{lammer03}).
Such escape could significantly modify or even entirely remove the primordial atmosphere of a closely orbiting rocky planet.

For M stars of both early \citep{shkolnik14} and mid--late \citep{schneider18} subtypes, average UV and X-ray emission stays elevated for at least 100~Myr after formation, followed by a decline.
Because UV and X-ray emission is magnetically controlled (even if not magnetically sourced; e.g., \citealt{hall08}), a reasonable expectation is that times of higher average emission will also correspond to greater rates of flares.
For white-light flares, this expectation is supported  by \citeauthor{davenport16a}'s (\citeyear{davenport16a}) recent flare analysis of lightcurves in the \textit{Kepler} archive.
This analysis revealed a $t^{-1}$ power-law decay in flare activity with Rossby number, implying that the stellar spin down that occurs with time drives declining flare rates.

This important result from \textit{Kepler} highlights the usefulness of the recent rise of staring observations, generally intended for detecting signatures of exoplanets, in constraining stellar flares.
\textit{Kepler} lightcurves were also analyzed by \cite{hawley14} for M dwarfs classified as active and inactive via H$\alpha$ emission, concluding that active M dwarfs flare more frequently than their inactive (but also earlier subtype) counterparts.
However, the distinction between groups in flare rates was less distinct than in H$\alpha$ equivalent widths.

\cite{davenport16b} used \textit{MOST} data to determine the flare frequency distribution (FFD, the relationship between the energy and occurrence rate of flares) for the host of the nearest habitable-zone planet, Proxima Centauri, showing that superflares ($E_{bol} > 10^{33}$~erg) could occur as often as 8~yr$^{-1}$.
Additional data from the \textit{Evryscope} time-domain survey recently detected one such flare on Proxima Centauri in 1344 h ($\approx2$~months) of data \citep{howard18}.
Assuming an associated particle event, \cite{howard18} concluded the planet, if bearing an Earth-like atmosphere, could have experienced substantial destruction of the ozone column.
What these optical and infrared photometry surveys detect is primarily the continuum (and ``conundruum;'' \citealt{kowalski13}) emission from the flares.
Continuum emission can vary considerably in temperature, as shown by \cite{kowalski13} in an analysis of \textit{U}-band spectra yielding blackbody fits spanning temperatures of 9,000--14,000~K.

Observations of flares at UV wavelengths directly affecting planetary atmospheres are more costly and accordingly rarer.
Until recently, they generally focused on specific objects with established track records of large flares, such as AD Leo (M4; e.g. \citealt{hawley03}).
However, \cite{welsh07} used \textit{GALEX} UV data to identify 52 flares among 49 stars and concluded that M0--M5 stars exhibit more energetic flares than M6--M8 types.

More recently, L18 identified flares in FUV \textit{Hubble Space Telescope (HST)} data collected for the MUSCLES Treasury Survey \citep{france16p}, using these data to constrain FUV flare rates on M dwarfs and comparing inactive and active stellar samples.
They found that FFDs between the two samples were nearly identical when flares were characterized by their equivalent duration rather than absolute energy.
Equivalent duration is a metric that physically represents the time the star would have to spend in quiescence to emit as much energy within the bandpass of interest as the flare alone.
In essence, it normalizes flare energies by the star's quiescent luminosity in the same band.
Hence, the consistency in M-dwarf FUV equivalent duration FFDs likely reflects a close correlation between the processes powering quiescent emission and flare emission.
This consistency is convenient for predicting flare rates where only a time-averaged measurement of a star's FUV emission is available.
However, L18 had no observations of young M stars to test if this consistency also spans M star ages.

Presently, the collection of a new treasury-scale dataset of FUV \textit{HST} observations is nearing completion as part of the HAbitable Zones and M dwarf Activity across Time (HAZMAT) program (PI Shkolnik).
Earlier phases of the HAZMAT program used archival \textit{GALEX} data to measure the evolution of X-ray, FUV, and NUV flux across time in early \citep{shkolnik14} and mid--late \citep{schneider18} M stars.
These data were also used to measure FUV and NUV variability of M stars, revealing greater overall variability in the FUV versus the NUV and an increase in NUV variability toward later spectral types \citep{miles17}.
The dedicated HAZMAT \textit{HST} program aims to gather spectra covering most of the FUV and NUV for groups of stars with three well-constrained ages: the Tucana-Horologium (Tuc-Hor) group at 40~Myr \citep{kraus14,bell15}, the Hyades cluster at 650~Myr \citep{maeder81,perryman98,martin18}, and field stars (several Gyr).
Observations of the Tuc-Hor moving group members have been completed, consisting of 12 M stars spanning types M0.0--M2.3.

In this work, we identified flares that occurred on these Tuc-Hor objects during the HAZMAT \textit{HST} observations.
One of these flares was a superflare ($E_\mathrm{bolo} >10^{33}$ erg, greater than any flare yet observed on the Sun), meriting more detailed scrutiny.
In Section \ref{sec:obs} we describe the observations and stellar sample.
In Section \ref{sec:analysis} we present the detected flares, power-law fits to the flare distributions, and flare rates tabulated for each star.
A discussion follows in Section \ref{sec:discussion} in which we address the implications of the flare distribution and compare to an equivalent dataset for field M dwarfs of slightly later spectral type (\ref{sec:ffds}), discuss the substantial quiescent variations in flux observed between flares (\ref{sec:qvars}), detail the superflare detected during the observations (\ref{sec:hazflare}), and explore the implications of such a flare for a planetary atmosphere (\ref{sec:planets}).
Results are summarized in Section \ref{sec:summary}.


\section{Observations}
\label{sec:obs}

\begin{deluxetable*}{lllrrrrrrr}
\tablewidth{0pt}
\tabletypesize{\footnotesize}
\rotate
\tablecaption{Relevant properties of the stars in the sample.\label{tbl:starprops}}
\tablehead{ \multicolumn{2}{c}{Star Name} & \colhead{Spectral} & \colhead{Distance\tablenotemark{b}} & \colhead{GALEX} & \colhead{GALEX} & \colhead{Total Exp.} & \colhead{Obs. Date} \\ \colhead{This Work} & \colhead{2MASS} & \colhead{Type\tablenotemark{a}} & \colhead{[pc]} & \colhead{NUV\tablenotemark{c}} & \colhead{FUV\tablenotemark{c}} & \colhead{Time\tablenotemark{d} [ks]} & \colhead{YYYY-MM-DD} }
\startdata
J03315 &  J03315564-4359135 & M0.0 & $ 45.275 \pm 0.071 $ & $ 18.311 \pm 0.039 $ & $ 20.49 \pm 0.19 $ & 10.12 & 2017-07-20\\
J00240 &  J00240899-6211042 & M0.2 & $ 44.2 \pm 1.1 $ & $ 18.498 \pm 0.056 $ & $ 20.11 \pm 0.19 $ & 1.64 & 2017-08-30\\
J02543 &  J02543316-5108313 & M1.1 & $ 43.76 \pm 0.21 $ & $ 19.192 \pm 0.064 $ & $ 20.65 \pm 0.18 $ & 1.42 & 2017-08-12\\
J00393 &  J00393579-3816584 & M1.4 & $ 40.241 \pm 0.070 $ & $ 18.494 \pm 0.045 $ & $ 20.42 \pm 0.17 $ & 1.26 & 2017-09-21\\
J23261 &  J23261069-7323498 & M1.5 & $ 46.294 \pm 0.059 $ & $ 18.762 \pm 0.070 $ & $ 20.18 \pm 0.19 $ & 1.35 & 2017-08-18\\
J01521 &  J01521830-5950168 & M1.6 & $ 39.765 \pm 0.040 $ & $ 19.005 \pm 0.054 $ & $ 20.37 \pm 0.15 $ & 1.28 & 2017-08-17\\
J22025 &  J22025453-6440441 & M1.8 & $ 43.705 \pm 0.098 $ & $ 19.208 \pm 0.060 $ & $ 20.88 \pm 0.19 $ & 1.33 & 2017-08-30\\
J02125 &  J02125819-5851182 & M1.9 & $ 48.061 \pm 0.049 $ & $ 19.251 \pm 0.067 $ & $ 21.04 \pm 0.26 $ & 1.84 & 2017-10-04\\
J02365 &  J02365171-5203036 & M2 & $ 38.847 \pm 0.051 $ & $ 18.379 \pm 0.038 $ & $ 20.49 \pm 0.18 $ & 9.96 & 2017-08-09\\
J02001 &  J02001277-0840516 & M2.1 & $ 36.926 \pm 0.069 $ & $ 19.038 \pm 0.021 $ & $ 20.601 \pm 0.050 $ & 1.76 & 2017-08-31\\
J22463 &  J22463471-7353504 & M2.3 & $ 50.224 \pm 0.074 $ & $ 19.657 \pm 0.025 $ & $ 21.54 \pm 0.22 $ & 2.30 & 2017-09-01\\
J23285 &  J23285763-6802338 & M2.3 & $ 46.048 \pm 0.050 $ & $ 19.124 \pm 0.035 $ & $ 20.48 \pm 0.15 $ & 1.26 & 2017-08-19\\
\enddata

\tablenotetext{a}{From \citep{kraus14} with the exception of J02365 \citep{torres06}. Errors are $\pm$ one subtype.}
\tablenotetext{b}{From \textit{GAIA} DR2 \citep{brown16a, brown18}.}
\tablenotetext{c}{For information on the GALEX magnitude system, see \citep{morrissey07}.}
\tablenotetext{d}{For exposures using G130M grating of COS only. Each observation consisted of four separate exposures.}

\end{deluxetable*}

The analysis presented here utilized data from 12 M-star members of the Tuc-Hor young moving group taken with the Cosmic Origins Spectrograph (COS), a photon-counting FUV and NUV spectrograph aboard \textit{HST}.
The data were collected specifically for the HAZMAT program (program ID 14784, PI Shkolnik).
The stars in the sample were identified as members of the Tuc-Hor association using 3D space velocities and youth indicators by \cite{shkolnik11,shkolnik12} and \cite{kraus14}.
This association is very young, yet old enough that the circumstellar disks have been dispersed, permitting unobstructed observations.
Thus data on this association is valuable to investigations of stellar evolution, such as studies of how stellar activity changes with age \citep{shkolnik14}.
The sample includes spectral types ranging from M0.0--M2.3.
This range mimics that of \cite{shkolnik14} for ease of comparison and ensures that stellar age and not spectral type is the primary independent variable sampled by this survey.

The HAZMAT sample selection process ensured that all stars had confirmed ages and no known visual and spectroscopic binaries.
In addition to 3D velocities consistent with Tuc-Hor membership, the age of the stars is supported by optical spectra that exhibit \Ha\ in emission without lithium in absorption (lithium is rapidly burned after about 10~Myr for late K and early M stars; \citealt{kraus14}).
Specifically, the lithium depletion boundary in Tuc-Hor objects implies an age of 35-45 Myr for the group, consistent with isochrone fits by \cite{bell15} yielding a prediction of $45\pm4$~Myr.
We adopt 40~Myr as the age of the group.
Table \ref{tbl:starprops} gives selected properties for each of the objects.

For comparison with earlier work, we restrict this analysis to the FUV data taken with COS's G130M grating.
The wavelength coverage of this configuration is $\sim$1170~--~1430~\AA\ and includes strong emission lines of \Cii, \Ciii, \Siiii, \Siiv, and \Nv\ formed in the stellar transition region and a smattering of weaker lines (including some coronal iron lines).
\lya\ and \Oi\ are in the bandpass also, but are typically lost to contamination by telluric emission (geocoronal airglow).
Because COS uses a photon-counting detector (the raw data being simply a list of detector position and time for each photon), the flux can be measured in arbitrary bins of wavelength and time to create lightcurves, integrated spectra, and subsampled spectra.
While there are instrumental limits to the wavelength and time resolution, in practice signal-to-noise ratio (S/N) requirements ultimately set much coarser resolution limits.

Most objects in this program were exposed for roughly 1~ks, with two exposed for 10~ks.
To mitigate the effect of fixed-pattern noise on integrated spectra, the observations are dithered, with gaps of about 100~s between each exposure.
For longer total exposures, these gaps instead last the $\sim$45~minute duration over which a target is obscured by the Earth during \textit{HST's} orbit.
These gaps can obscure portions of flares and introduce ambiguity into whether a single event or multiple events occurred and the total energy of the event(s).
The uncertainty this introduces in deriving characteristics of the overall flare population is further addressed in Section \ref{sec:analysis}.


\section{Analysis}
\label{sec:analysis}

We identified flares in the data using the FLAIIL (FLAre Identification in Interrupted Lightcurves\footnote{\url{https://www.github.com/parkus/flaiil}}) algorithm described by L18, including the same $\mathrm{FUV}_{130}$ bandpass.
Briefly, the method estimates quiescent fluxes using a Gaussian Process model with a covariance kernel of the form $\sigma_x^2 e^{-\Delta t / \tau}$, where $\Delta t$ is the difference in time between data points and $\sigma_x^2$ and $\tau$ are parameters specifying the variance and decorrelation timescale of the data.
The $x$ in $\sigma_x^2$ denotes that the variations being fit are those in excess of what would be expected from measurement uncertainty.
A penalty is applied for power at 0.1~Hz to mitigate overfitting of noise and flares and a flat line is used when the likelihood ratio of such a model to the Gaussian process is $<$2.

Following quiescence fitting, the lightcurve is divided into ``runs'' of points above and below quiescence, and runs with an integrated area 5$\sigma$ above quiescence are flagged as flares and those with $3\sigma$ above or below quiescence are flagged as suspect.
The process is iterated, with each iteration conducting a maximum likelihood fit of the Gaussian Process model to the nonflare and nonsuspect points.
Iteration is terminated when the same points are successively flagged as flare or suspect.

A bonus of the identification procedure is that the Gaussian Process fit provides a measure of the amplitude and timescale of the star's quiescent variations.
For the young M stars, quiescent variations can be quite significant (Section \ref{sec:qvars}).
A useful relationship for predicting the expected sample standard deviation, $[S]$, as a function of lightcurve binning, $\Delta t$, based on the Gaussian Process fit is
\begin{equation}
[S_x] = \left(\frac{2\sigma_x^2}{(\Delta t/\tau)^2} \left(\Delta t/\tau + e^{-\Delta t/\tau} - 1\right)\right)^{1/2}.
\end{equation}
In addition to quantifying quiescent variations via a Gaussian Process, we compute the 60~s ``excess noise'' metric as per \cite{loyd14} and the relative median absolute deviation (MAD) as per \cite{miles17} to facilitate comparisons between these works.
The results are tabulated in Table \ref{tbl:qvars}.

To produce lightcurves, we integrated nearly the same wavelength range as the FUV$_{130}$ band from L18.
This spans most of the COS~G130M bandpass, excluding areas contaminated by telluric emission, roughly covering 1170--1270 and 1330--1430~\AA.
We record the total energy, $E$; equivalent duration, $\delta$; peak flux, $F_p$; full width at half max, FWHM; rise time; and decay time for each flare.
Rise times, decay times, and the FWHM are made ambiguous by noise fluctuations and complex, multi-peaked structures in some flares.
Thus, the rise and decay times are measured as the time between last crossing of the quiescent level and first crossing of half-max at the flare start and vice versa at the flare end.
The FWHM is the sum of all time during which the flare's flux was above half-max.

\begin{deluxetable*}{lrrrrrrrrl}
\tablewidth{0pt}
\tabletypesize{\scriptsize}
\tablecaption{Identified flares. \label{tbl:flares}}
\tablehead{ \colhead{Star} & \colhead{$\delta$} & \colhead{$E$} & \colhead{$t_\mathrm{peak}$} & \colhead{$F_{\mathrm{peak}}$} & \colhead{$\frac{F_\mathrm{peak}}{F_q}$\tablenotemark{a}} & \colhead{Rise Time} & \colhead{FWHM} & \colhead{Decay Time} & \colhead{Complex?\tablenotemark{b}} \\ \colhead{} & \colhead{s} & \colhead{$10^{30}$ erg} & \colhead{MJD} & \colhead{$10^{-13}\ \frac{\mathrm{erg}}{\mathrm{cm}^2\mathrm{s\ \AA}}$} & \colhead{} & \colhead{s} & \colhead{s} & \colhead{s} & \colhead{} }
\startdata
J02365 & $ 6736 \pm 40 $ & $ 130.64 \pm 0.55 $ & 57974.5621 & $ 66.7 \pm 6.8 $ & $ 63.1 \pm 6.7 $ & 150 & 55 & 770 & Y\\
 & $ 832 \pm 14 $ & $ 16.13 \pm 0.26 $ & 57974.3864 & $ 6.82 \pm 0.80 $ & $ 7.35 \pm 0.89 $ & 88 & 37 & 330 & Y\\
 & $ 359.7 \pm 9.8 $ & $ 6.98 \pm 0.19 $ & 57974.4511 & $ 4.98 \pm 0.59 $ & $ 5.64 \pm 0.69 $ & 17 & 65 & 180 & N\\
 & $ 48.2 \pm 6.1 $ & $ 0.93 \pm 0.12 $ & 57974.4311 & $ 1.72 \pm 0.22 $ & $ 2.60 \pm 0.34 $ & 22 & 53 & 14 & \nodata\\
J01521 & $ 1961 \pm 45 $ & $ 12.78 \pm 0.19 $ & 57982.9865 & $ 4.43 \pm 0.53 $ & $ 13.9 \pm 2.7 $ & 130 & 74\tablenotemark{c} & \nodata & \nodata\\
J03315 & $ 405 \pm 14 $ & $ 7.37 \pm 0.26 $ & 57954.8492 & $ 2.68 \pm 0.31 $ & $ 4.65 \pm 0.55 $ & 150 & 23 & 75 & \nodata\\
 & $ 101.6 \pm 8.4 $ & $ 2.00 \pm 0.17 $ & 57954.9052 & $ 1.49 \pm 0.18 $ & $ 2.85 \pm 0.34 $ & 16 & 53 & 41 & \nodata\\
 & $ 46.0 \pm 7.1 $ & $ 0.77 \pm 0.12 $ & 57954.7906 & $ 1.02 \pm 0.12 $ & $ 2.49 \pm 0.30 $ & 3.9 & 93 & 5.1 & \nodata\\
 & $ 40.4 \pm 6.2 $ & $ 0.77 \pm 0.12 $ & 57954.8642 & $ 1.58 \pm 0.17 $ & $ 3.05 \pm 0.34 $ & 7.1 & 9.4 & 25 & \nodata\\
J22025 & $ 300 \pm 14 $ & $ 6.53 \pm 0.21 $ & 57995.2908 & $ 15.1 \pm 1.6 $ & $ 16.8 \pm 2.6 $ & 19 & 9.7 & 110 & N\\
J02543 & $ 214 \pm 10 $ & $ 2.26 \pm 0.11 $ & 57977.8361 & $ 1.55 \pm 0.20 $ & $ 4.37 \pm 0.58 $ & 26 & \nodata\tablenotemark{c} & \nodata & \nodata\\
J00240 & $ 96.7 \pm 9.0 $ & $ 1.53 \pm 0.14 $ & 57995.8298 & $ 1.99 \pm 0.25 $ & $ 3.93 \pm 0.67 $ & 29 & 8.1 & 7.7 & \nodata\\
J23285 & $ 89.9 \pm 9.8 $ & $ 1.23 \pm 0.13 $ & 57984.0811 & $ 1.07 \pm 0.14 $ & $ 2.98 \pm 0.49 $ & \nodata & 56\tablenotemark{c} & 12 & \nodata\\
J02001 & $ 66.4 \pm 6.3 $ & $ 0.752 \pm 0.071 $ & 57996.5731 & $ 2.14 \pm 0.24 $ & $ 4.08 \pm 0.48 $ & 26 & \nodata\tablenotemark{c} & \nodata & \nodata\\
 & $ 42.0 \pm 6.0 $ & $ 0.480 \pm 0.069 $ & 57996.5586 & $ 1.15 \pm 0.15 $ & $ 2.64 \pm 0.35 $ & 9.8 & 52 & 12 & \nodata\\
 & $ 31.7 \pm 4.1 $ & $ 0.356 \pm 0.046 $ & 57996.5666 & $ 1.45 \pm 0.17 $ & $ 3.11 \pm 0.36 $ & 51 & \nodata\tablenotemark{c} & \nodata & \nodata\\
 & $ 21.5 \pm 4.5 $ & $ 0.296 \pm 0.062 $ & 57996.5068 & $ 1.46 \pm 0.17 $ & $ 2.73 \pm 0.33 $ & 18 & 17 & 9.6 & \nodata\\
J00393 & $ 40.7 \pm 5.1 $ & $ 0.404 \pm 0.051 $ & 58017.9703 & $ 1.39 \pm 0.16 $ & $ 3.72 \pm 0.43 $ & 39 & 15\tablenotemark{c} & \nodata & \nodata\\
\enddata

\tablenotetext{a}{Ratio of peak flux to quiescent flux.}
\tablenotetext{b}{Subjective determination of the complexity of the flare shape based on its deviation from an impulse-decay, generally due to multiple peaks. No data indicates the flare was not well-enough resolved or the classification was particularly ambiguous.}
\tablenotetext{c}{Flare cut off by the start or end of an exposure.}

\tablecomments{Uncertainties are statistical and do not reflect systematic effects due to choices made in the flare identification and measurement algorithm. See text for a discussion of the effect of these choices.}
\end{deluxetable*}

\begin{deluxetable*}{lrrrrr}
\tablewidth{0pt}
\tabletypesize{\scriptsize}
\tablecaption{Quiescence fit parameters and measurements of quiescent variability. \label{tbl:qvars}}
\tablehead{\colhead{Star} & \colhead{$\sigma_{x, \mathrm{GP}}$\tablenotemark{a}} & \colhead{$\tau_\mathrm{GP}$\tablenotemark{a}} & \colhead{$\sigma_{x, \mathrm{LF14}}$\tablenotemark{b}} & \colhead{MAD$_\mathrm{rel}$\tablenotemark{c}} \\ \colhead{} & \colhead{} & \colhead{s} & \colhead{} & \colhead{}}
\startdata
J03315 & $0.049_{-0.003}^{+0.154}$ & $19306_{-2252}^{+301823}$ & $0.112_{-0.012}^{+0.014}$ & $ 0.1090 \pm 0.0082 $\\
J00240 & $0.37_{-0.05}^{+1.74}$ & $18709_{-2385}^{+346318}$ & $0.40_{-0.09}^{+0.14}$ & $ 0.199 \pm 0.056 $\\
J02543 & $0.130_{-0.008}^{+0.517}$ & $16689_{-1825}^{+357107}$ & $0.26_{-0.07}^{+0.13}$ & $ 0.242 \pm 0.051 $\\
J00393 & $0.061_{-0.005}^{+0.269}$ & $10330_{-1644}^{+277252}$ & $0.194_{-0.051}^{+0.084}$ & $ 0.114 \pm 0.026 $\\
J23261 & $<5.5\times10^{-5}$ & \nodata & $0.104_{-0.026}^{+0.038}$ & $ 0.046 \pm 0.014 $\\
J01521 & $0.155_{-0.037}^{+0.033}$ & \nodata & $<0.61$ & $ 1.31 \pm 0.22 $\\
J22025 & $0.46_{-0.03}^{+2.96}$ & $2013_{-225}^{+193452}$ & $0.53_{-0.13}^{+0.24}$ & $ 0.286 \pm 0.099 $\\
J02125 & $0.052_{-0.005}^{+0.233}$ & $11384_{-1802}^{+373590}$ & $0.184_{-0.043}^{+0.070}$ & $ 0.090 \pm 0.019 $\\
J02365 & $0.030_{-0.019}^{+0.010}$ & \nodata & $0.065_{-0.011}^{+0.012}$ & $ 0.747 \pm 0.012 $\\
J02001 & $0.095_{-0.010}^{+0.404}$ & $15512_{-2725}^{+369976}$ & $0.18_{-0.07}^{+0.14}$ & $ 0.118 \pm 0.030 $\\
J22463 & $0.104_{-0.007}^{+0.474}$ & $6357_{-694}^{+272637}$ & $0.251_{-0.052}^{+0.065}$ & $ 0.144 \pm 0.035 $\\
J23285 & $0.101_{-0.034}^{+0.022}$ & \nodata & $0.151_{-0.048}^{+0.084}$ & $ 0.095 \pm 0.038 $\\
\enddata

\tablenotetext{a}{Pertains to covariance kernel function, $\sigma_x^2 e^{-\Delta t / \tau}$, of the Guassian Process used to model quiescent variations. Values and uncertainties are based on the \nth{16}, \nth{50}, \nth{84} percentiles of the MCMC samples.}
\tablenotetext{b}{``Excess noise'' at 60 s cadence per \cite{loyd14}. Values and uncertainties are based on the \nth{16}, \nth{50}, \nth{84} percentiles of the analytical solution  of the posterior distribution.}
\tablenotetext{c}{Median Absolute Deviation relative to median per \cite{miles17}. Uncertainties are based on the \nth{16}, \nth{50}, \nth{84} percentiles from bootstrapped samples. Uses a 100~s cadence and includes flares.}

\end{deluxetable*}

\begin{deluxetable*}{lrrrrrrrrr}
\rotate
\tablewidth{0pt}
\tabletypesize{\scriptsize}
\tablecaption{Properties derived from the constrained flare distribution in equivalent duration (see text for details). \label{tbl:pew_stats}}
\tablehead{\colhead{Star} & \colhead{$N_\mathrm{fit}$\tablenotemark{a}} & \colhead{$N_\mathrm{all}$\tablenotemark{b}} & \colhead{$\log\left(\nu(>10 \mathrm{\ s})\right)$} & \colhead{$\log\left(\nu(>10^3 \mathrm{\ s})\right)$} & \colhead{$\log\left(\nu(>10^6 \mathrm{\ s})\right)$} & \colhead{$\log\left(E_3/E_q\right)$\tablenotemark{c}} & \colhead{$\log\left(E_6/E_q\right)$\tablenotemark{c}} & \colhead{$\log\left(\delta_\mathrm{crit}\right)$\tablenotemark{d}} & \colhead{$\delta_\mathrm{min}$\tablenotemark{e}} \\ \colhead{} & \colhead{} & \colhead{} & \colhead{$\log(\mathrm{d}^{-1})$} & \colhead{$\log(\mathrm{d}^{-1})$} & \colhead{$\log(\mathrm{d}^{-1})$} & \colhead{} & \colhead{} & \colhead{$\log(\mathrm{s})$} & \colhead{s}}
\startdata
J03315 & 4 & 4 & $1.64_{-0.22}^{+0.26}$ & $0.41_{-0.33}^{+0.40}$ & $-1.39_{-0.69}^{+0.87}$ & $-1.41_{-0.21}^{+0.26}$ & $-0.15_{-0.48}^{+0.55}$ & $5.7_{-0.7}^{+5.8}$ & 17\\
J00240 & 1 & 1 & $1.76_{-0.42}^{+0.60}$ & $0.54_{-0.50}^{+0.66}$ & $-1.26_{-0.78}^{+0.96}$ & $-1.28_{-0.42}^{+0.60}$ & $-0.02_{-0.60}^{+0.74}$ & $5.2_{-1.0}^{+6.6}$ & 17\\
J02543 & 1 & 1 & $1.87_{-0.43}^{+0.61}$ & $0.64_{-0.49}^{+0.63}$ & $-1.19_{-0.78}^{+0.90}$ & $-1.18_{-0.42}^{+0.60}$ & $0.06_{-0.59}^{+0.70}$ & $5.2_{-1.1}^{+5.5}$ & 24\\
J00393 & 1 & 1 & $1.94_{-0.42}^{+0.61}$ & $0.71_{-0.48}^{+0.64}$ & $-1.11_{-0.77}^{+0.90}$ & $-1.10_{-0.42}^{+0.61}$ & $0.14_{-0.58}^{+0.71}$ & $5.2_{-1.4}^{+5.1}$ & 27\\
J23261 & 0 & 0 & $<2.5$ & $<1.3$ & $<-0.22$ & $-1.25_{-0.42}^{+0.60}$ & $-0.18_{-0.63}^{+0.72}$ & $5.7_{-1.4}^{+8.3}$ & 19\\
J01521 & 1 & 1 & $2.01_{-0.43}^{+0.62}$ & $0.79_{-0.48}^{+0.63}$ & $-1.03_{-0.75}^{+0.89}$ & $-1.03_{-0.43}^{+0.61}$ & $0.22_{-0.58}^{+0.69}$ & $4.7_{-1.0}^{+5.1}$ & 37\\
J22025 & 1 & 1 & $2.12_{-0.44}^{+0.60}$ & $0.91_{-0.47}^{+0.63}$ & $-0.91_{-0.75}^{+0.90}$ & $-0.92_{-0.42}^{+0.60}$ & $0.34_{-0.56}^{+0.70}$ & $4.4_{-1.0}^{+4.7}$ & 49\\
J02125 & 0 & 0 & $<2.5$ & $<1.2$ & $<-0.29$ & $-1.25_{-0.42}^{+0.59}$ & $-0.22_{-0.62}^{+0.70}$ & $5.4_{-1.1}^{+8.8}$ & 29\\
J02365 & 4 & 4 & $1.89_{-0.23}^{+0.27}$ & $0.69_{-0.30}^{+0.33}$ & $-1.07_{-0.64}^{+0.75}$ & $-1.15_{-0.21}^{+0.26}$ & $0.14_{-0.43}^{+0.47}$ & $5.1_{-0.6}^{+3.6}$ & 47\\
J02001 & 4 & 4 & $2.41_{-0.21}^{+0.24}$ & $1.21_{-0.31}^{+0.36}$ & $-0.56_{-0.65}^{+0.76}$ & $-0.63_{-0.21}^{+0.24}$ & $0.66_{-0.44}^{+0.50}$ & $4.1_{-0.6}^{+2.1}$ & 18\\
J22463 & 0 & 0 & $<2.7$ & $<1.3$ & $<-0.25$ & $-1.10_{-0.42}^{+0.62}$ & $-0.08_{-0.58}^{+0.68}$ & $5.1_{-1.2}^{+7.8}$ & 61\\
J23285 & 0 & 1 & $1.90_{-0.43}^{+0.61}$ & $0.56_{-0.50}^{+0.66}$ & $-1.44_{-0.83}^{+1.00}$ & $-1.17_{-0.43}^{+0.61}$ & $-0.08_{-0.61}^{+0.73}$ & $5.2_{-1.2}^{+7.6}$ & 22\\
All & 17 & 18 & $1.87_{-0.12}^{+0.13}$ & $0.68_{-0.23}^{+0.27}$ & $-1.09_{-0.59}^{+0.73}$ & $-1.17_{-0.11}^{+0.12}$ & $0.13_{-0.37}^{+0.44}$ & $5.3_{-0.5}^{+3.7}$ & \nodata\\
\enddata

\tablenotetext{a}{Number of flares used in the FFD fit, i.e. only those with equivalent durations where the survey was deemed sufficiently complete.}
\tablenotetext{b}{Total number of flares detected. If $N_\mathrm{all} > N_\mathrm{fit}$, then the difference represents flares not used in the FFD fits because they had equivalent durations below the threshold where the survey was deemed sufficiently complete.}
\tablenotetext{c}{Ratio of flare to quiescent energy emitted averaged over very long timescales based on the power-law fit, integrating flare energies from 10~--~$10^3$~s or 10~---~$10^6$~s.}
\tablenotetext{d}{Critical equivalent duration beyond which, if the power-law model holds, energy emitted by flares over long timescales will exceed the integrated quiescent emission. Error bars are defined by the location of the 5th and 95th percentiles.}
\tablenotetext{e}{Detection limit of each dataset.}

\end{deluxetable*}

\begin{deluxetable*}{lrrrrrrrrr}
\rotate
\tablewidth{0pt}
\tabletypesize{\scriptsize}
\tablecaption{Properties derived from the constrained flare distribution in absolute energy (see text for details). \label{tbl:energy_stats}}
\tablehead{\colhead{Star} & \colhead{$N_\mathrm{fit}$\tablenotemark{a}} & \colhead{$N_\mathrm{all}$\tablenotemark{b}} & \colhead{$\log\left(\nu(>10^{27} \mathrm{erg})\right)$} & \colhead{$\log\left(\nu(>10^{30} \mathrm{erg})\right)$} & \colhead{$\log\left(\nu(>10^{33} \mathrm{erg})\right)$} & \colhead{$E_\mathrm{min}$\tablenotemark{c}} \\ \colhead{} & \colhead{} & \colhead{} & \colhead{$\log(\mathrm{d}^{-1})$} & \colhead{$\log(\mathrm{d}^{-1})$} & \colhead{$\log(\mathrm{d}^{-1})$} & \colhead{10$^{27}$ erg}}
\startdata
J03315 & 4 & 4 & $3.01_{-0.49}^{+0.43}$ & $1.18_{-0.23}^{+0.27}$ & $-0.69_{-0.53}^{+0.63}$ & 3.2$\times10^{29}$\\
J00240 & 1 & 1 & $3.18_{-0.58}^{+0.65}$ & $1.38_{-0.42}^{+0.60}$ & $-0.48_{-0.63}^{+0.77}$ & 4.2$\times10^{29}$\\
J02543 & 1 & 1 & $3.09_{-0.58}^{+0.68}$ & $1.29_{-0.43}^{+0.62}$ & $-0.58_{-0.67}^{+0.80}$ & 2.9$\times10^{29}$\\
J00393 & 1 & 1 & $3.14_{-0.57}^{+0.67}$ & $1.33_{-0.42}^{+0.59}$ & $-0.55_{-0.66}^{+0.77}$ & 2.6$\times10^{29}$\\
J23261 & 0 & 0 & $<4.7$ & $<1.3$ & $<-2.0$ & 3.1$\times10^{29}$\\
J01521 & 1 & 1 & $3.13_{-0.57}^{+0.65}$ & $1.33_{-0.44}^{+0.60}$ & $-0.53_{-0.67}^{+0.81}$ & 2.4$\times10^{29}$\\
J22025 & 1 & 1 & $3.46_{-0.63}^{+0.70}$ & $1.67_{-0.43}^{+0.61}$ & $-0.21_{-0.61}^{+0.75}$ & 9.3$\times10^{29}$\\
J02125 & 0 & 0 & $<4.5$ & $<1.1$ & $<-2.2$ & 3.0$\times10^{29}$\\
J02365 & 4 & 4 & $3.25_{-0.52}^{+0.48}$ & $1.47_{-0.22}^{+0.26}$ & $-0.36_{-0.50}^{+0.55}$ & 9.2$\times10^{29}$\\
J02001 & 4 & 4 & $3.66_{-0.41}^{+0.40}$ & $1.85_{-0.23}^{+0.27}$ & $0.02_{-0.53}^{+0.60}$ & 2.2$\times10^{29}$\\
J22463 & 0 & 0 & $<4.6$ & $<1.5$ & $<-0.57$ & 4.6$\times10^{29}$\\
J23285 & 0 & 1 & $3.3_{-1.0}^{+1.4}$ & $0.0_{-1.1}^{+1.8}$ & $-3.5_{-1.4}^{+2.0}$ & 3.0$\times10^{29}$\\
All & 17 & 18 & $3.20_{-0.41}^{+0.36}$ & $1.38_{-0.12}^{+0.13}$ & $-0.44_{-0.45}^{+0.53}$ & \nodata\\
\enddata

\tablenotetext{a}{Number of flares used in the FFD fit, i.e. only those with equivalent durations where the survey was deemed sufficiently complete.}
\tablenotetext{b}{Total number of flares detected. If $N_\mathrm{all} > N_\mathrm{fit}$, then the difference represents flares not used in the FFD fits because they had equivalent durations below the threshold where the survey was deemed sufficiently complete.}
\tablenotetext{c}{Detection limit of each dataset.}

\end{deluxetable*}

We detected a total of 18 flares in 35.5~ks of exposure of 12 targets with COS~G130M.
Included in this sample is a flare that emitted $10^{32.1}$ erg in the FUV, exceeding the most energetic M-star flare previously observed in the FUV with \textit{HST} by about a factor of 30.
Because these stars are comparatively distant (38--53~pc) relative to the older, less FUV-luminous M dwarfs previously observed by \textit{HST}, the smallest detectable flares are correspondingly more energetic.
However, their range of equivalent durations, 20--6700~s, is similar to the previously observed flares.
The detected flares and their properties are tabulated in Table \ref{tbl:flares}.

For each star, as well as the aggregated sample, we fit power laws to the FFDs.
Preferences in the literature vary in exactly how to specify this power law, but in this work we will use the cumulative form
\begin{equation}
\nu = \mu E^{-\alpha},
\end{equation}
where $\nu$ is the frequency of flares with energy (equivalent duration) greater than $E$ ($\delta$), $\mu$ specifies the rate of flares with unit energy (equivalent duration), and $\alpha$ is the index of the power law.

We fit FFDs in both absolute energy and equivalent duration using the same method used by L18.
The method computes a likelihood of the individual event energies (i.e., events are not binned) given a power-law index as well as the Poisson-likelihood of the number of events observed once the rate constant is applied.
The MCMC sampler \texttt{emcee} \citep{foreman13} is then used to sample the posterior of these parameters.
Tables \ref{tbl:pew_stats} and \ref{tbl:energy_stats} give the parameters of these fits for each star and the flares aggregated from all stars.

These fits then enable estimates of the rate of flares above various energy and equivalent duration limits.
Considering long timescales over which many flares occur, the FFD fits allow predictions of the cumulative energy emitted by flares and the ratio of this energy to that emitted by quiescence.
Further, one can estimate the ``critical equivalent duration,'' defined as the limit to which integrating the FFD fit predicts energy emitted by the star's flares will exceed that emitted by the star in quiescence (L18).
We used the MCMC samples of the fit parameters to sample the posterior distribution of these derived quantities, thereby accounting for the strong correlation between the fit parameters.

No single star exhibited enough flares ($\gtrsim5$) to effectively constrain the index, $\alpha$, of the power law in a fit to the FFD.
However, by applying an a priori constraint on this index, the rate constant, $\mu$, of flares can also be constrained.
This in turn provides constraints on flare rates and other derived quantities.
Hence, we use this technique for the individual stars.
For the necessary constraint on the power-law index, we use the posterior probability distribution from the fit to the aggregated flares from all stars as a prior for each individual star.
The fits to the aggregated flares are
\begin{equation}
\nu(E) = 10^{1.38_{-0.13}^{+0.12}}\ \mathrm{d}^{-1}\ \left(\frac{E}{10^{30}~\mathrm{erg}}\right)^{-0.61_{-0.13}^{+0.15}}
\label{eq:FFDabs}
\end{equation}
and
\begin{equation}
\nu(\delta) = 10^{0.68_{-0.27}^{+0.23}}\ \mathrm{d}^{-1}\ \left(\frac{\delta}{1000~\mathrm{s}}\right)^{-0.59_{-0.13}^{+0.15}}.
\end{equation}
Using the stabilized Kolmogorov-Smirnov test recommended by \citep{maschberger09}, we find that both power laws provide acceptable fits to the data, with $p$-values of 0.6 (energy) and 0.7 (equivalent duration).

It is critical to note that the uncertainties specified throughout this work are statistical.
However, systematics, such as the exposure gaps mentioned in Section \ref{sec:obs}, and subjective choices made in constructing the flare identification algorithm influence the results.
These affect the overall number and measured characteristics (most importantly energy) of the identified flares.
We adjusted the parameters of our flare identification and FFD fitting apparatus within reasonable ranges and observed changes to assess the degree of the effect on the FFD power-law index.
This includes the readiness with which events closely spaced in time relative to the overall duration are associated, including those broken up by an exposure gap.
The appendices of L18 include a more detailed discussion of the various parameters.
For this work, we found that the total number of flares varied between 16 and 22 and the index of the equivalent duration FFD fit varied between 0.4 and 0.8 according to our analysis choices.
This is important to bear in mind when interpreting FFDs and derived values.

\begin{figure*}
\includegraphics{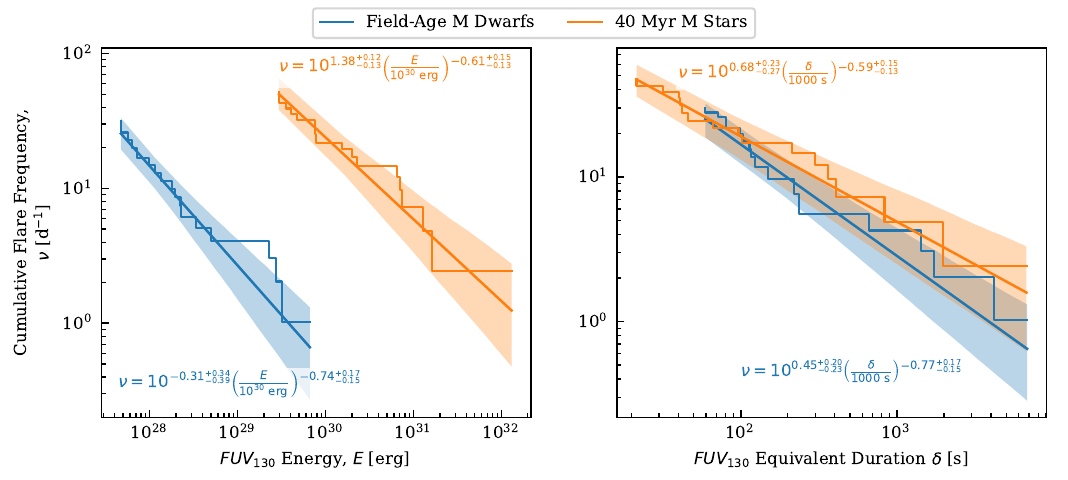}
\caption{Comparison of the flare frequency distributions (FFDs) of field-age (from L18) and 40~Myr M stars. Stepped lines give the cumulative flare rate (corrected for differing detection limits between observations), straight lines are power-law fits, and shaded regions are 1$\sigma$ errors on those fits. In absolute energy (left), the limiting energy for flares occurring less than 5 times per day is 100--1000$\times$ greater on the 40~Myr versus field-age stars. However, in equivalent duration (right) the rates are statistically indistinguishable over the observed range. Note that spectral types differ somewhat, spanning M0.0--M2.3 for the 40~Myr sample and M1.5--M3.5 for the field sample.}
\label{fig:ffds}
\end{figure*}


\section{Discussion}
\label{sec:discussion}

\subsection{M Star Flares at 40 Myr are 100--1000$\times$ More Energetic than at Field Ages}
\label{sec:ffds}
The primary question driving this analysis was ``how does FUV flare activity on M stars evolve over their lifetimes?''
This is answered in Figure \ref{fig:ffds}, in which the FFDs of the 40 Myr stars and the field-age stars are compared, using both equivalent duration and absolute energy.
The field-age curve is taken from L18 and relies on observations acquired by the MUSCLES program \citep{france16} of GJ~667C (M1.5), GJ~436 (M2.5), GJ~832 (M2/3), GJ~1214 (M3), GJ~581 (M3), and GJ~876 (M3.5), with ages ranging from 1-9~Gyr \citep{anglada13, torres08, sanz10, berta11, selsis07, rivera10}.

The energy above which flares occur less than 5 times per day is 100--1000$\times$ greater for the 40 Myr versus the field-age M stars.
Comparing, instead, the rates of flares with energies $>10^{30}$ erg, those on 40~Myr stars occur 20-100$\times$ more frequently.
(Note that quiescent flux over the course of the flare is subtracted when computing energies, so  the difference in the absolute energy FFDs is not due simply to integrating different quiescent luminosities.)
The FFD of the active star sample from L18 (not plotted to avoid clutter) falls between these two.

The two distributions overlap extensively when flares are characterized by their equivalent duration.
This suggests that the conclusion of L18 that all M dwarfs flare similarly in the FUV when flares are characterized by their equivalent duration can be extended to young M stars as well.
The difference in the power-law indices implies that this will not hold true beyond the plotted range, but this difference is not statistically significant.
Any true difference in the indices and thus rate of higher-energy flares, if it exists, can only be resolved with more variability monitoring of young M stars in the FUV.
The similarity of the distributions in equivalent duration means the differences in absolute energy are due almost entirely to the differing quiescent FUV luminosity of the stars.

The stellar samples differ somewhat in their makeup of spectral types. The 40~Myr stars vary from M0.0 to M2.3 while the field-age stars vary from M1.5 to M3.5.
The only field-age object with a spectral type confidently within the range of 40~Myr sample is GJ~667C, and the rates of $\delta=1000$~s and $E=10^{30}$~erg flares estimated for this star fall squarely within the rest of the field-age sample.
If anything, the differences in spectral types would be expected to lessen the gap in the absolute energy FFDs due to greater flare activity on later-type M dwarfs \citep{hilton11} and create a gap in the equivalent duration FFDs due to differing stellar conditions.
Hence, the conclusions regarding the difference in absolute flare energies and similarities in equivalent duration are likely to hold for samples more consistent in spectral type.
Upcoming observations for the HAZMAT program will add three field-age M stars with M0.5, M0.5, and M2.0 types, enabling a future flare analysis of a field-age sample with stellar types in closer alignment with the other age groups sampled by HAZMAT.

The differences in flare activity accord well with the measurements made by \cite{shkolnik14} and \cite{schneider18} of changes in FUV and NUV activity over time.
Their measurements used observations from \textit{GALEX} whose FUV bandpass spans 1340--1800~\AA\ (containing the strong emission lines \Siiv, \Civ, and \Heii), while the present analysis relies on emission from an 1170--1270~+~1330--1430~\AA\ range (containing \Ciii, \Siiii, \Nv, \Cii, \Siiv).
\cite{shkolnik14} measure a median \textit{GALEX} FUV flux for the 40~Myr objects in their sample that is 20$\times$ that of the field objects, similar to the drop in the rate of $>10^{30}$~erg FUV$_{130}$ flares measured in this analysis.
In other words, it appears quiescent flux levels and the energy output of flares drop in synchrony, further supporting the consistency  of M-star  equivalent-duration FFDs found by  L18.
If this similarity between evolution in quiescent flux levels and flare rates holds, then the 650~Myr (Hyades) stars will flare at a slightly reduced (factor of a few) rate relative to the 40~Myr (Tuc-Hor) stars.
\textit{HST} observations of the Hyades cluster for the HAZMAT program will be complete this year.

The uniformity of M dwarf flares in equivalent duration seems a natural consequence of magnetic activity both heating the stellar transition region during quiescence as well as producing flares.
It is evidence against other mechanisms of quiescent heating, such as upwelling shocks \citep{hall08}, unless somehow these shocks are linked to flares.
This relationship should not to be taken to imply quiescent emission is due simply to unresolved flares.
As L18 noted, extrapolating the FFD of M dwarfs to zero and integrating could only account for a small fraction of quiescent emission (though this is not true for flares in some specific emission lines like \Siiv).
This means that flare and quiescent FUV emission either arise from distinct mechanisms or the FFDs steepen considerably (and consistently) at unresolved flare energies.
Consistency in equivalent duration FFDs does not appear to extend to the Sun, which exhibits FFDs with rates three orders of magnitude below those of M dwarfs in analogous EUV emission (L18).

Regardless of the source of the consistency in equivalent duration FUV FFDs for M stars, it is a convenient fact.
It implies models of stellar flaring, such as those that might be used in assessing the impacts to planetary atmospheres, can utilize an estimate of a star's FUV luminosity as a single parameter to describe the star's FUV flare activity within the range of energies thus far observed.

An important feature of FUV flares on M stars is that they might dominate the energy budget of FUV emission from such stars over timescales long enough to include rare, highly energetic flares not yet observed in the FUV.
For field age objects, if the power-law FFD estimated by L18 extends 2--5 orders of magnitude beyond the most energetic flare identified in that analysis, then flare emission begins to dominate.
Considering the most energetic flare here observed was 30$\times$ more energetic, the likelihood that flares indeed dominate over quiescence is increased.
For the distribution of aggregated 40~Myr M star flares, this limit is $10^{5}$--$10^{9}$~s in equivalent duration ($10^{33}$--$10^{37}$~erg), 1--5 orders of magnitude beyond the most energetic event observed.


\subsection{Substantial Quiescent Variability in Young Ms}
\label{sec:qvars}

\begin{figure}
\plotone{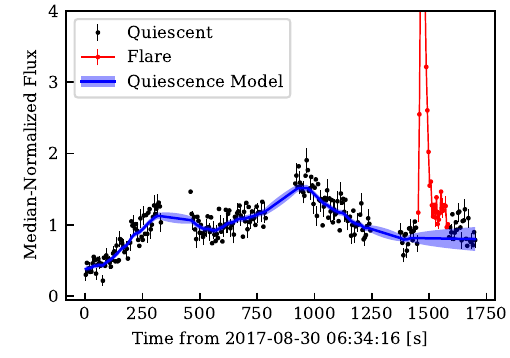}
\caption{$\mathrm{FUV}_{130}$ lightcurve of J22025 showing the factor of a few variations during quiescence.}
\label{fig:qvar_example}
\end{figure}

FUV emission from the young M stars in this sample exhibits substantial levels of quiescent variability.
As part of the analysis, we computed several measures of variability, presented in Table \ref{tbl:qvars}.
An example of these variations are those of J22025 where the ratio of the maximum to the minimum observed flux during quiescence is 6.
The $\mathrm{FUV}_{130}$ lightcurve of this star is depicted in Figure \ref{fig:qvar_example}.
Although the quiescent variations of J22025 are among the largest in the sample, the  values in Table \ref{tbl:qvars} make it clear that significant quiescent  variations are normal for the observed stars.

The FUV variability of M stars has also  been surveyed by \cite{loyd14} and \cite{miles17}.
\cite{loyd14} conducted an analysis of archival \textit{HST} data similar to that conducted here, but using flux from specific emission lines that  are major contributors to  the flux in the band employed for the  present analysis (\Cii, \Siiii, \Siiv).
The most comparable stars  in  their sample are GJ~832 (M2/3) and AU~Mic (M1), a known young flare star that is still contracting toward the main sequence.
GJ~832 exhibits ``excess noise,'' $\sigma_x$, of 0.1--0.15 in the  major lines, similar to the lower end of the present sample.
AU Mic exhibits excess noise $<0.03-0.1$, below all  but two of the stars in the present sample.

\cite{miles17} analyzed archival \textit{GALEX} data using the MAD relative to the median, \madrel.
Only a few points were available to characterize variability, precluding the identification and removal of flares in that work.
Therefore, \madrel\ measurements for the 40~Myr M stars in Table \ref{tbl:qvars} include flares, though the statistic is mostly insensitive to them.
The 40~Myr M stars exhibit \madrel\ values within the upper half of the \cite{miles17} sample, with J01521 and J22025 reaching the upper extreme.

In some cases, the variations exhibit a clear timescale.
This is quantified by the decay-timescale for autocorrelations in the Gaussian Process we used to model quiescence.
Because a penalty was applied for nonsmoothness (to avoid overfitting noise and potentially ``fitting out'' flares), decay timescales are systematically longer than otherwise.
Timescales range from tens of minutes to hours, with J22025 showing the clearest timescale by eye to its variations.
Lifting the nonsmoothness penalty for that object results in a time-constant of $1800_{-1000}^{+4000}$~s for the Gaussian Process fit to quiescence.
This is too rapid to be attributed to stellar rotation, but matches the expected timescale of convective granulation \citep{kjeldsen95}.
Therefore, we posit that convective motions are modulating the magnetic heating in one to a few localized active regions in this star and perhaps others exhibiting large-amplitude variations with a clear timescale.
The amplitude of the variations suggests a limit to the number of active regions that could be contributing significantly to the quiescent flux.
A greater number of independently evolving regions would result in those fluctuations averaging out (unless the fluctuations are of correspondingly larger amplitude).
A simple test of this hypothesis is to compare simultaneous optical and FUV observations.
If convection is driving quiescent FUV variability, the timescale of optical and FUV fluctuations for a given star should match.

It might seem likely that stars with larger-amplitude quiescent variations would have more tumultuous magnetic heating leading to more flares.
Yet the stars exhibiting flares in this sample do not exhibit anomalously high quiescent variability.
Alternatively, suppressed quiescent variations might indicate that magnetic energy that would have powered them is instead building toward a catastrophic release in the form of a flare, i.e. that smaller-amplitude quiescent variations would be accompanied by more frequent flares.
Yet this too is not apparent in the data.
Indeed, the stars exhibiting flares span nearly the full range of quiescent variability.
Future datasets providing longer baselines, particularly from staring NUV and FUV observations by the upcoming \textit{SPARCS} \citep{shkolnik16} and \textit{CUTE} \citep{fleming18} cubesats, could determine whether quiescent variations and flares are truly uncorrelated.

\begin{figure}
\plotone{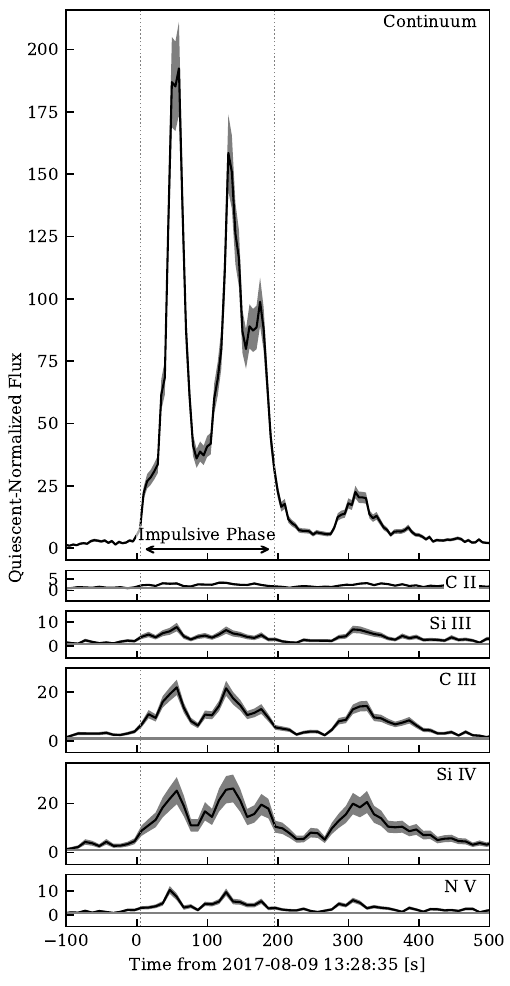}
\caption{Lightcurves of major emission lines and continuum regions for the Hazflare normalized by their quiescent flux.
Dotted lines bracket the impulsive phase that was integrated to produce the spectrum in Figure \ref{fig:hazflare_spectra}.
Sizes of the axes have been adjusted such that the scales are closely similar.
All lines are the same as those identified in Figure \ref{fig:hazflare_spectra} with multiple components coadded, except that \Ciii\ refers only to the multiplet at 1175~\AA, not the line at 1247~\AA.
Horizontal lines at unity are shown to guide the eye.}
\label{fig:hazflare_lightcurves_normed}
\end{figure}

\begin{figure}
\plotone{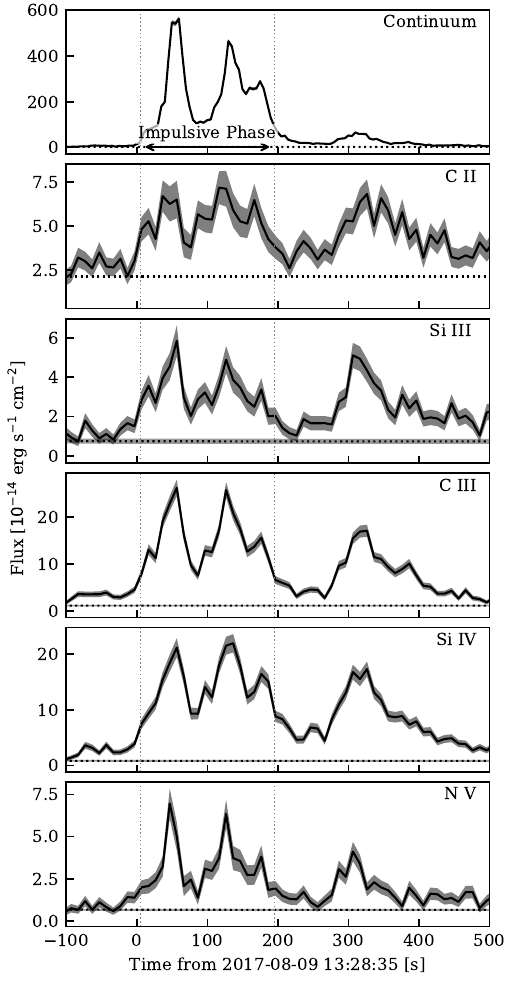}
\caption{Lightcurves of major emission lines and continuum regions for the Hazflare in absolute flux.
The best-fit continuum and uncertainty is shown as the dotted horizontal line.
Dotted vertical lines bracket the impulsive phase that was integrated to produce the spectrum in Figure \ref{fig:hazflare_spectra}.}
\label{fig:hazflare_lightcurves_absolute}
\end{figure}


\subsection{\emph{Hubble} FUV Spectrophotometry of an M Star Superflare }
\label{sec:hazflare}
\begin{figure*}
\plotone{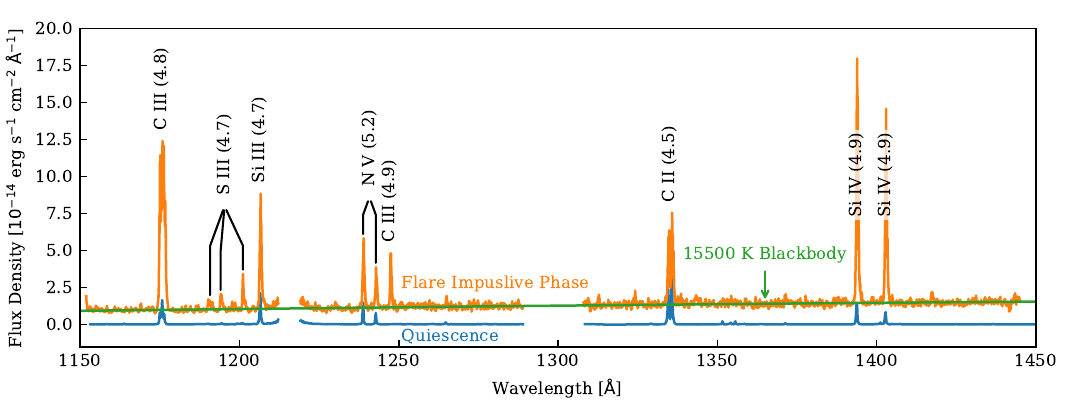}
\caption{The spectrum of the Hazflare averaged over its impulsive phase (orange; Figure \ref{fig:hazflare_lightcurves_normed} shows integration limits) with the best-fit blackbody overplotted (green) and compared to the quiescent spectrum of the star (blue). Major lines are identified with the log formation temperature (in Kelvin) from a model solar atmosphere \citep{dere09} shown in parenthesis. Regions within the COS detector gap and regions contaminated by geocoronal emission have been removed.}
\label{fig:hazflare_spectra}
\end{figure*}

\begin{figure}
\plotone{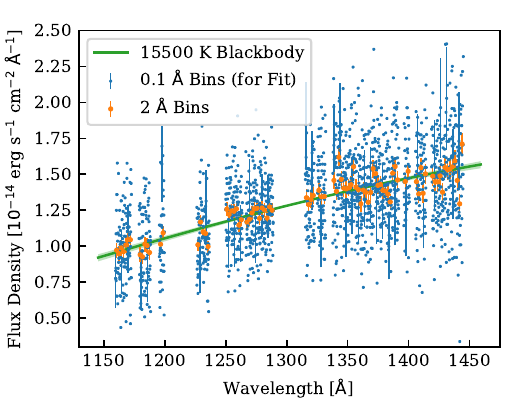}
\caption{Blackbody fit to the continuum regions of the Hazflare spectrum during its impulsive phase.}
\label{fig:hazflare_bb_fit}
\end{figure}

The most energetic flare observed on the Tuc-Hor objects was a superflare \citep{schaefer00}, meaning its bolometric energy exceeded $10^{33}$ erg, more than any recorded solar flare.
We have dubbed this event the Hazflare.
The Hazflare occurred on the M2.0 object GSC~8056-0482 (J02365 in this work), peaking at 2017 August 09 13:29 UT and is, to our knowledge, the most energetic stellar flare yet observed in the FUV by \textit{HST}.
Lightcurves and spectra for this flare are plotted in Figures \ref{fig:hazflare_lightcurves_normed}, \ref{fig:hazflare_lightcurves_absolute}, and \ref{fig:hazflare_spectra}.
Confident classification of this event as a superflare is possible due to a well-resolved blackbody continuum in the FUV that can be extrapolated across all wavelengths as a lower limit on the bolometric flare energy.
A fit to this continuum, shown in Figure \ref{fig:hazflare_bb_fit}, implies the blackbody had a characteristic temperature of $15,500\pm400$~K across the impulsive phase of the flare and alone emitted a wavelength-integrated energy of $10^{33.44\pm0.04}$~erg over the entire flare.
In the \textit{U} band, commonly used for ground-based flare observations, the blackbody emitted $10^{32.47\pm0.05}$ erg.
Including an appropriately scaled version of the fiducial flare energy budget in L18, which includes estimates of nonthermal emission in the unobserved FUV and inobservable EUV, increases the bolometric energy estimate to $10^{33.6_{-0.2}^{+0.1}}$~erg (where we have assumed an order-of-magnitude uncertainty in the EUV scalings).

This observation is of particular value because superflares are common on stars (e.g., \citealt{davenport16a}), yet spectrophotometry of such flares in the UV, the band most relevant to planetary atmospheric photochemistry, is rare.
Superflares are estimated from \textit{Kepler} data to occur on M0--M4 dwarfs at a frequency of a few per day \citep{yang17}. 
Photochemical models exploring the effects of flares on planetary atmospheres have thus far relied primarily on observations of the 1985 Great Flare on AD Leo \citep{hawley91,segura10,tilley17}, a flare estimated to emit a bolometric energy of $10^{34}$ erg.
The Great Flare also showed a clear continuum in FUV emission, and overall the continuum was responsible for at least an order of magnitude more overall energy emitted by the flare than lines, consistent with the Hazflare.
However, the Great Flare observations, made with the \textit{International Ultraviolet Explorer}, saturated in the strongest emission lines, degrading their accuracy.
The present observation clearly resolved the temporal evolution in all major emission lines over the course of the flare, except for \lya\ and \Oi\ since these are contaminated by Earth's geocoronal emission.

\begin{figure*}
\plotone{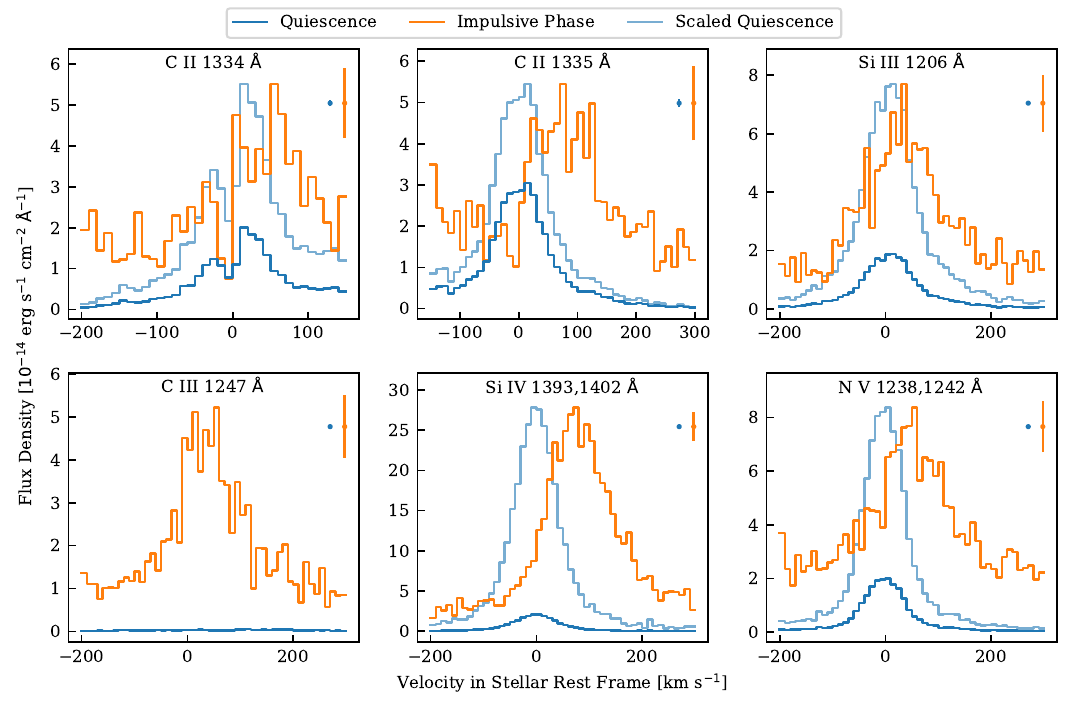}
\caption{Major emission lines during the Hazflare versus quiescence. Dark blue shows the quiescent line profile, orange the profile during the impulsive phase of the Hazflare, and light blue is a scaled quiescent profile to facilitate the comparison of line shapes (not shown for \Ciii\ since no line is resolved in quiescence). The \Nv\ and \Siiv\ doublets are coadded, but this is not done for \Cii\ due to ISM absorption affecting one component. The spectra are binned to 10~km~\pers\ resolution, roughly the absolute wavelength accuracy of COS. Bars representative of the error at the line peaks are shown on the top right of each panel.}
\label{fig:hazflare_lines}
\end{figure*}

Major emission lines during the Hazflare are shown in Figure \ref{fig:hazflare_lines}.
The lines show redshifts from 50--80~km~\pers\, similar to previous observations of FUV lines during flares (e.g., \citealt{hawley03}, L18), that signify downward flows of material during the flare.
During the flare, a strong line appears at 1247~\AA\ of which there is no hint in the star's quiescent spectrum.
Based on the CHIANTI solar atmosphere model \citep{dere09}, we identify this line to be a transition of \Ciii.

The \Ciii~1247~\AA\ line is dipole allowed, with a similar transition probability to the components of the \Ciii\ complex at 1175~\AA\ (not shown in Figure \ref{fig:hazflare_lines} because of blending).
However, the upper level of the 1247~\AA\ transition is 5.6~eV more energetic, hence it is more difficult to populate.
The distinct on/off nature of this line between the flaring and quiescent state could be an important constraint on the physical conditions elicited by the flare.
A possible explanation is that the upper level of the 1247~\AA\ transition is populated by collisional excitation from the upper level of the 1175~\AA\ transitions.
Because radiative de-excitation to produce the 1175~\AA\ lines is fast, a threshold rate of collisions would need to be reached to excite to the upper level of the 1247~\AA\ transition.
However, such excitation might then quench the 1175~\AA\ line, and this is not observed.
We encourage further exploration of this topic in future modeling work.

The spectral energy budget of the Hazflare is within the scatter in energy budgets of  M~dwarf flares of lower energy as analyzed by L18.
The most important feature, energetically, of the flare spectrum is the blackbody.
Multiwavelength observations of AD Leo flares by \cite{hawley03} that all exhibited roughly 9000~K emission yielded ratios of blackbody emission  to \Siiv\ emission of 100--200.
In comparison, this ratio for the 15,500~K Hazflare is 240.
This could be an important constraint for simulations of flares, as it suggests the partitioning of energy between FUV and blackbody emission remains relatively constant even though one might reasonably expect the relative contribution of the blackbody to be a factor of $(15,000/9,000)^4 \approx 8$ higher in a 15,000~K versus a 9,000~K flare.

The Hazflare is not of exceptional FUV energy when normalized by the quiescent luminosity of the star, i.e., when energy is quantified as equivalent duration.
Several \textit{HST} observations of other M dwarfs yielded flares of greater FUV equivalent duration, specifically those of 15 ks and 12 ks on Prox~Cen and 6.8 ks on GJ~876 (L18).
The temporal evolution of the Hazflare is similar to these events, exhibiting multiple rapid increases and drops in emission, the ``impulsive phase'' of the flare.
(Note some other authors reserve the use of the term ``impulsive'' for the initial flare rise, e.g., \citealt{bookbinder92}.)

This structure is also qualitatively mimicked in the integrated \textit{U}-band flux of the Great AD Leo Flare.
However, the impulsive phase of the Great AD Leo Flare lasts about 4$\times$ longer than the Hazflare.
If the two flares have similar continuum emission, this alone could potentially explain the greater energy of the Great Flare.
\cite{hawley91} took the continuum radiation of the Great AD Leo Flare to be that of recombination continua and unresolved lines and did not estimate blackbody temperatures or filling factors.

In determining the temperature of the Hazflare blackbody, we addressed the effect of extinction by the ISM.
Extinction is much stronger at FUV wavelengths than the \textit{U} band wavelengths where blackbody emission is usually resolved, and the 41.7~pc distance to this star could potentially contain sufficient dust for significant absorption.
Using the 3D local ISM dust model, we estimated a worst-case $E(B-V)$ extinction of 0.015 mag.
As a worst-case $R_V$ (a parameter that sets the relationship between extinction and wavelength), we take a value of 2 given at least one known sight line has an $R_V$ near this extreme: 2.1 for HD 210121 \citep{welty92}.
This yielded a predicted absorption of 10\%--7\% across G130M bandpass and increased the temperature of the best-fit blackbody by 300 K.
We considered this effect minimal and this scenario unlikely, so we neglected reddening in the remainder of our analysis.

The 15,500~K temperature of the Hazflare blackbody is somewhat uncommon, above the 9,000--14,000~K range observed by \cite{kowalski13} (though they estimate up to $\sim$1000~K uncertainties).
However, \textit{GALEX} observations of a flare on GJ~3685A (M4) suggest a blackbody temperature of 50,000~K based on the ratio of broadband FUV to NUV flux \citep{robinson05} and blue-optical observations of continuum emission from flares on YZ~CMi (also M4) could be consistent with plasma temperatures as high as 170,000~K \citep{kowalski13,kowalski18}.
Flare modeling assuming heating  by a beam of nonthermal electrons directed at the stellar surface by \cite{kowalski15} has successfully reproduced blackbody-like emission $\approx$10,000~K in the blue optical.
Earlier models using lower electron beam energy fluxes  could not  reproduce this emission, suggesting that further increasing the energy flux in such models might reproduce the Hazflare continuum emission.
It is noteworthy that the the continuum shape in these models, though it is blackbody-like, is more attributable to changing optical depth with wavelength than the temperature of an optically thick source.

To the knowledge of the authors, the Hazflare is the first event in which a blackbody emission temperature could be constrained using FUV spectra.
It is possible even higher temperature emission is present in some flares, but is not well-constrained by \textit{U} band spectra since they are further within the Rayleigh-Jeans tail of the Planck curve.
Future flare observations in the FUV and NUV could reveal more events like the Hazflare or hotter.

\begin{figure}
\plotone{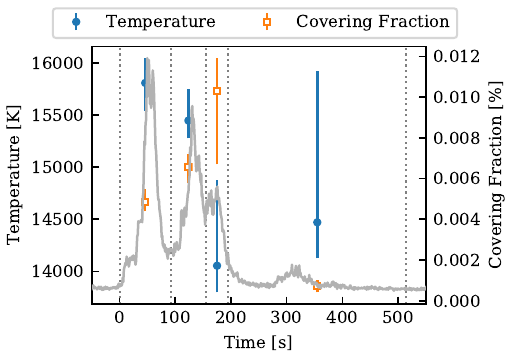}
\caption{Evolution of blackbody emission over the course of the Hazflare. Dashed vertical lines delineate the integration ranges used to produce spectra for Planck-function fits. The lightcurve of emission in the continuum bands is plotted as well for reference.}
\label{fig:bb_evol}
\end{figure}

High blackbody temperatures were sustained throughout the Hazflare.
In Figure \ref{fig:bb_evol}, we show the temperature of the blackbody in three separate time intervals (limited signal precludes further divisions in time).
The flare peaks initially nearer to 16,000~K and decays to 14,000~K.
We plot the filling factor of the flare in Figure \ref{fig:bb_evol} as well, a quantity specifying the area of blackbody emission required to yield the flare continuum flux as a fraction of the star's visible hemisphere.
In comparison to the AD Leo observations of \cite{hawley03,kowalski13}, the high temperature of the Hazflare yields a comparatively small filling factor despite a larger flare energy.
Along with the hotter flares characterized in \cite{kowalski13}, the Hazflare lends support to the idea that heating area, duration, and intensity are all important in determining the total energy emitted by a flare, rather than only area and duration.

During the Hazflare, the covering fraction increases with each successive peak in the triple-peaked flare, yet the successive drops in blackbody temperature result in a lower energy emitted by each peak.
Similar behavior was present in a flare on the M dwarf YZ~CMi observed by \cite{mochnacki80}.
However, evolution of the blackbody during flares does not seem to follow a predictable pattern.
Temperature and covering fraction evolved in lock-step during a YZ~CMi flare observed by \cite{kowalski13}.
In several flares on the M dwarf AD~Leo observed by \cite{hawley03}, temperature and covering fraction sometimes track well, and sometimes do not.
When they do not, emission closely tracks covering fraction rather than temperature.
For the Hazflare temperature and covering fraction do not track, and temperature appears to better account for changes in emission.


\subsection{Planetary Implications of the Hazflare}
\label{sec:planets}

Flares like the Hazflare will bombard orbiting planets, influencing the immediate state as well as long-term evolution of their atmospheres.
For the young M2.0 star that produced the Hazflare, the eventual main-sequence HZ lies 0.1--0.2~AU from the star (adopting a stellar mass of 0.4~$M_\sun$; \citealt{kopparapu13}).
Note that a planet orbiting in this range will be too hot to support liquid water currently because the star is about twice as luminous at its present 40~Myr age as it will be several Gyr in the future, i.e. the HZ will move inward by about 40\% as the star evolves onto the main sequence, according to the evolutionary tracks of \cite{baraffe15}.
The decline in luminosity that drives the HZ evolution is predicted to last until 250~Myr, then over the next 10~Gyr the star's luminosity is predicted to increase by 9\% .

Because the 9,000--14,000~K range of flare blackbodies thus far observed peak in the NUV, they are well-suited to photolyzing ozone.
Ozone has a peak in its photolysis cross section at 2550~\AA, corresponding to an 11,400~K blackbody.
Varying the blackbody temperature of equal-energy flares from 5,000--15,000~K can change the photolysis rate ($J$-value) of unshielded \OIII\ molecules,  by a factor of a few (L18).
During the impulsive phase of the Hazflare, the blackbody emission could drive \JOIII$\ = 0.04$~\pers\ for a planet at mid-HZ, about 5$\times$ that of the Sun at Earth \citep{loyd16}.

The recombination of O and \OII\ is very fast, as is the thermal dissociation of \OIII.
In consequence, O and \OIII\ are rapidly exchanged, quickly reaching an equilibrium ratio under a given set of conditions.
This ratio is driven by the NUV radiation field, so as the NUV flux of a flare evolves, the \OIII\ column evolves almost in lock-step.
Ultimately, the dissociation of \OII\ by the flare's FUV radiation will provide additional atomic oxygen reservoir that, once the elevated NUV flux abates, yields an \OIII\ column that is greater than at the flare onset.
This phenomenon is clear in the simulations of \cite{segura10} and \cite{tilley17} when only EM radiation from flares is considered.
If energetic particles are assumed to accompany a flare, they produce \OIII-destroying catalysts that then drive a dramatic depletion of \OIII\ over timescales well beyond the duration of any single flare.
For the Hazflare, applying the solar scaling of \cite{youngblood17p} yields an estimate for the fluence of $>$10~MeV protons in the main-sequence HZ of $10^{5.6\pm 0.8}$~proton~cm$^{-2}$~s$^{-1}$~sr$^{-1}$, two orders of magnitude above the largest solar observation given in \cite{youngblood17p}.

Of course, an atmosphere need not be Earth-like, and flares could be an important photolyzer of other molecules.
Noteworthy is the methane prevalent in Titan's atmosphere whose photolysis products go on to assemble long hydrocarbon chains that contribute to an atmospheric haze (see \citealt{horst17} for a recent review).
As with ozone, secondary catalytic reactions are important here, as C$_2$H$_2$ and C$_4$H$_2$ further destroy \CHIV\ once they are produced from its photolysis products.
Similarly, work by \cite{hu13} has shown that in reducing atmospheres, photolysis of surface-outgassed H$_2$S and SO$_2$ species can yield S and S$_2$ that then polymerize into hazes.
In these instances, flares like the Hazflare could be an important additional source term for producing haze-forming monomers.
Hazes are important to exoplanet observations because they could obscure absorption features in transmission spectroscopy (e.g., \citealt{kreidberg14,kawashima18}).
They are also important to life, as they could shield the surface from UV radiation and might also act as a source of biological precursor molecules \citep{hu13}.

Similar to ozone, the photolysis cross section of H$_2$S has a peak in the NUV. 
However, photolysis cross sections for \CHIV\ are skewed more toward the FUV.
Thus differences in the blackbody temperature of a flare can have a more dramatic effect on this molecule (L18).
During the impulsive phase of the flare, the Hazflare blackbody alone could drive \JCHIV$\ = 0.01$~\pers\ at mid-HZ, 1600$\times$ that of the Sun at 1 AU.

Another relevant feature of the Hazflare's 15,500 K blackbody is emission in the EUV at wavelengths where photons can ionize H.
Such emission would heat the upper atmospheres of orbiting planets, potentially powering additional thermal atmospheric escape from the planet.
The blackbody emits an energy in the 100--912~\AA\ EUV range equivalent to 2\% of that given  by the fiducial flare of L18 scaled to match the Hazflare's \Siiv\ emission.
The contribution of EUV energy to the fiducial flare is based on scaling from solar data (see L18 for details).

Energy conservation requires that the EUV (blackbody and otherwise) irradiation of the Hazflare could not have removed more than $\sim$10$^{12}$~g of atmospheric mass from an Earth-gravity planet at mid-HZ, $\sim$10$^{-9}$ the mass of Earth's atmosphere.
The fact that the relatively brief (35~ks) cumulative exposure of the HAZMAT campaign captured such a flare suggests that they are common.
These observations constrain the rate of flares emitting $>10^{33}$~erg in the FUV (i.e., FUV superflares) to 0.1--1~d$^{-1}$ (Table \ref{tbl:energy_stats}).
If the mass-loss efficiency for flares is similar to the canonical value of 0.1 for steady-state flux (e.g., \citealt{murray09}), then the accumulated atmospheric ``erosion'' by such flares could be significant over timescales of hundreds of Myr.
\cite{chadney17} modeled the effects of a flare on mass loss from hot Jupiters and found that it could not explain variations seen in the \lya\ transit of HD~189733b.
More modeling is needed to determine the efficiency of flare EUV emission in removing atmospheric mass over a range of planetary parameters (e.g. mass), physical regimes of mass loss, and flare energies.
In addition, particle events associated with M star flares are a persistent unknown.
If they accompany highly energetic events as on the Sun such as predicted by the scaling relations of \cite{youngblood17p}, then flare erosion of planetary atmospheres would be more severe.


\section{Summary and Future Work}
\label{sec:summary}
With the aim of constraining the FUV flare activity of young M dwarfs, we identified and analyzed FUV flares that occurred on a sample of 12 M0--M2.3 stars in the Tuc-Hor association, age 40~Myr, during 35.5 ks of \textit{HST} COS-G130M observations.
We identified 18 flares in total and fit a power law to the distribution of these flares.

These young M stars are indeed more active in terms of flares than a comparison sample of older, inactive field M1.5--M3.5 dwarfs for which an identical flare analysis was carried out by L18.
Specifically, on the 40~Myr M stars flares with energy $>10^{30}$~erg occur 20-100$\times$ more frequently than on the field-age M stars.
Alternatively, the limiting energy for flares occurring at a rate of $<5$~d$^{-1}$ is 100-1000$\times$ greater.
This elevation in flare activity at young ages mimics that of the average emission in the GALEX FUV and NUV bands  \citep{shkolnik14,schneider18}.

When the flare distributions are specified in equivalent duration rather than absolute energy, a metric that normalizes the flare energy by the stars' quiescent luminosities in the same band, we find the distributions of the 40~Myr and field-age M stars closely overlap.
This complements the finding of L18 that active versus inactive M stars show no significant difference in FUV flare activity when equivalent duration is used to characterize their flares, extending it to old versus young samples as well.
In addition, the power-law fit to the equivalent duration distribution implies that more overall energy will be emitted in the FUV from flares versus quiescence if that power-law extends another 1--5 orders of magnitude beyond the most energetic flare in the sample.

The most energetic flare that occurred during the observations was a superflare (bolometric energy $>10^{33}$ erg).
A strong lower limit can be placed on the bolometric flux of this flare because of the greatly elevated continuum emission manifested by the flare, nearly 200$\times$ quiescent levels.
This made the continuum bright enough for its slope and curvature to be clearly resolved, permitting a blackbody fit that implies $15,500\pm400$~K emission was responsible for this continuum emission over the flare.
Stitching together the measured emission, blackbody fit, and the fiducial flare template of L18, we estimate a bolometric energy of this flare of $10^{33.6_{-0.2}^{+0.1}}$~erg, approaching the estimated $10^{34}$~erg of the 1985 Great Flare on AD Leo.
The hot blackbody emission of this flare would be a powerful photolyzer of most molecules in planetary atmospheres due to the high broadband FUV flux.

This and previous UV flare analyses beg further study and observations in several key areas.
Whether flares actually dominate quiescent emission is a question that cannot be confidently resolved until there are observations in the UV of sufficient cumulative time to constrain the high-energy tail of the FFD.
If the FFD for FUV flares on 40~My M stars obeys Eqn.~\ref{eq:FFDabs} out to $10^{36}$~erg flares (close to the maximum observed by \textit{Kepler}; \citealt{yang17}), these stars would need to be observed for several months to obtain meaningful constraints on the rate of such flares.
Additional observations are needed to fully characterize the diversity of continuum flux among flares, as the blackbody temperature of this emission is critical to photochemical models.
Further modeling work is needed in the area of flare erosion of planetary atmospheres to asses the (in)significance of this erosion in driving the lifelong evolution of planetary atmospheres.
Finally, diagnostics of particle events associated with stellar flares are imperative, as particles likely have much more severe implications for planetary atmospheres than flare EM radiation, yet their severity and even existence in relation to M-star flares has yet to be strongly observationally constrained.

\acknowledgments
R.O.P.L. and E.S. gratefully acknowledge support from NASA \textit{HST} Grant  HST-GO-14784.001-A for this work. R.O.P.L. thanks Brittany Miles, Adam Kowalski, Wilson Cauley, Melodie Kao, Tyler Richey-Yowell, Ella Osby, and Allison Youngblood for helpful discussions relating to this work.

\facilities{\textit{Hubble Space Telescope}}

\software{\texttt{extinction} (v0.3.0, \citealt{barbary16}, \url{http://extinction.readthedocs.io/})
\texttt{emcee} (v2.2.0, \citealt{foreman13}, \url{http://dfm.io/emcee/current/})
\texttt{celerite} (v0.3.0, \citealt{foreman17}, \url{http://celerite.readthedocs.io})
\texttt{FLAIIL} (\citealt{loyd18}, \url{https://github.com/parkus/flaiil}),
\texttt{FFD} (\citealt{loyd18}, \url{https://github.com/parkus/ffd})}

\end{document}